\documentclass[12pt]{article}
\usepackage{amscd,amssymb,amsmath,latexsym}
\usepackage[mathscr]{euscript}
\usepackage{epsfig}
\title{Correlations Estimates in the Lattice Anderson Model\footnote{Authors
supported in part by NSF grants DMS 06009565
(JVB), 0503784 (PDH), and 0245210 (GS).}
\vspace{.5cm}}

\author{Jean V.\ Bellissard$^{1}$, Peter D.\ Hislop$^{2}$, G\"unter
Stolz$^{3}$\\
{\small $^1$ Georgia Institute of Technology, School of Mathematics,
Atlanta GA 30332-0160}\\
{\small $^2$ Department of Mathematics, University of Kentucky,
Lexington, KY  40506-0027}\\
{\small $^3$ Department of Mathematics, UAB,
Birmingham AL 35294-1170}
\vspace{.2cm}}

\date{ }

\newtheorem{theo}{Theorem}

\newtheorem{lemma}{Lemma}
\newtheorem{coro}{Corollary}

\newcommand{\beq}{\begin{equation}}
\newcommand{\eeq}{\end{equation}}
\newcommand{\ba}{\begin{array}}
\newcommand{\ea}{\end{array}}
\newcommand{\bea}{\begin{eqnarray}}
\newcommand{\eea}{\end{eqnarray}}

\newcommand{\Hh}{{\mathcal H}}

\newcommand{\id}{{\mathbf 1}}

\newcommand{\CM}{{\mathbb C}}
\newcommand{\EM}{{\mathbb E}}

\newcommand{\NM}{{\mathbb N}}

\newcommand{\PM}{{\mathbb P}}
\newcommand{\RM}{{\mathbb R}}

\newcommand{\ZM}{{\mathbb Z}}

\newcommand{\uV}{\vec{u}}
\newcommand{\vV}{\vec{v}}

\newcommand{\TR}{{\rm Tr}}                         

\begin{document}

\maketitle

\begin{abstract}
We give a new proof of correlation estimates for arbitrary moments
of the resolvent of random Schr\"odinger operators on the lattice
that generalizes and extends the correlation estimate of Minami for
the second moment. We apply this moment bound to obtain a new
$n$-level Wegner-type estimate that measures eigenvalue correlations
through an upper bound on the probability that a local Hamiltonian
has at least $n$ eigenvalues in a given energy interval. Another
consequence of the correlation estimates is that the results on the
Poisson statistics of energy level spacing and the simplicity of the
eigenvalues in the strong localization regime hold for a wide class
of translation-invariant,
selfadjoint, lattice operators with decaying off-diagonal terms
and random potentials.
\end{abstract}



\vspace{.5cm}


\vspace{.2cm}


\section{Introduction: Correlation Estimates and Energy Level Statistics}

Correlations between various families of random variables associated
with disordered systems are an important aspect governing the
transport properties of the system. For example, the conductivity is
expressible in terms of the second moment of the one-electron
spectral density. Another example is the correlation between the
energy levels of noninteracting electrons for finite-volume systems
and their behavior in the thermodynamic limit. Some of the first
studies of energy level correlations were made by Molchanov
\cite{[Molchanov]} and by Minami \cite{Mi96} for systems in the
strong localization regime. Molchanov \cite{[Molchanov]} studied a
family of random Schr\"odinger operators in one-dimension with a
random potential given by $q(t, \omega) = F(x_t(\omega))$, where
$x_t(\omega)$ is Brownian motion on a compact manifold $K$ and $F$
is a smooth, real-valued, Morse function on $K$. It is known that
this model exhibits Anderson localization at all energies (cf.\
\cite{[PF],[CL]}). Minami \cite{Mi96} studied the lattice Anderson
model (see (\ref{minami05.eq-andersonmodel00})) in any dimension with
a bounded random Anderson-type potential for energies in the strong
localization regime. These authors proved that, under certain
hypotheses, the normalized distribution of electron energy levels in
the thermodynamic limit is Poissonian. This is interpreted to mean
that there is no level repulsion (nor attraction) between energy
levels in the thermodynamic limit provided the energy lies in the
strong localization region. This is in contrast to the expected
behavior when the energy lies in the region of transport and strong
correlations between energy levels are expected. In this case,
the expected eigenvalue spacing
distribution is a Wigner-Dyson distribution (cf.\ \cite{[SSSLS]}).

The precise formulation of this result is as follows. The standard
Anderson model studied by Minami is given by the following random
Hamiltonian acting on $\ell^2(\ZM^d)$,

\begin{equation}
\label{minami05.eq-andersonmodel00}
H_\omega \psi(x) \;=\;
\sum_{y;|y-x|=1} \;
  \psi(y)
  \;+\;
   V (x)\psi(x), ~~~x \in \ZM^d,
\hspace{2cm}
  \psi \in \ell^2(\ZM^d)\,,
\end{equation}

\noindent where the potential $\omega = \left(
V(x)\right)_{x\in\ZM^d}$ is a family of independent, identically
distributed random variables with common distribution with a density
$\rho(V (0))$ such that $\|\rho\|_{\infty} = \sup_{V(0)} \rho(V(0))
< \infty$. Let $H_\Lambda$ denote the restriction of $H_\omega$ to a
box $\Lambda \subset \ZM^d$ with Dirichlet boundary conditions. The
spectrum of $H_\Lambda$ is finite discrete and the eigenvalues
$E_j^\Lambda (\omega)$ are random variables. For any subset $J
\subset \RM$, we let $E_\Lambda (J)$ be the spectral projection for
$H_\Lambda$ and $J$. The integrated density of states (IDS) $N(E)$
is defined by \beq\label{ids1} N(E) = \lim_{|\Lambda| \rightarrow
\infty} \frac{ \EM (Tr E_\Lambda ( (-\infty, E])) }{ |\Lambda|} ,
\eeq when this limit exists. It is known (cf.\ \cite{[CL],[PF]})
that, for the lattice models considered here, this function exists
and is Lipschitz continuous (at least if the density has compact
support). Consequently, it is almost everywhere differentiable, and
its derivative $n(E)$ is the density of states (DOS) at energy $E$.
In order to describe energy level correlations, we focus on the
spectrum near a fixed energy $E$. We define a point process $\xi (
\Lambda;E) (dx) $ by \beq\label{poisson1} \xi ( \Lambda; E) (dx) =
\sum_{j \in \NM} ~\delta(( | \Lambda| (E_j^\Lambda (\omega) - E) - x
) ~dx \eeq The rescaling by the volume $|\Lambda|$ reflects the fact
that the average eigenvalue spacing is proportional to $|
\Lambda|^{-1}$. Minami proved the following theorem.

\begin{theo}\label{minamitheorem1}
Consider the standard Anderson model
(\ref{minami05.eq-andersonmodel00}) and suppose the DOS $n(E)$
exists at energy $E$ and is positive. Suppose also that the
expectation of some fractional moment of the finite-volume Green's
function decays exponentially fast as described in (\ref{expect1}).
Then, the point process (\ref{poisson1}) converges weakly as
$|\Lambda| \to \infty$ to the Poisson point process with intensity
measure $n(E) ~dx$.
\end{theo}

Minami's result requires two technical hypotheses: 1) the density of
states $n (E)$ must be non-vanishing at the energy $E$ considered,
and 2) the expectation of some fractional moment of the
finite-volume Green's function decays exponentially.
Wegner\cite{[Wegner]} presented an argument for the nonvanishing of
the DOS $n(E)$ and a strictly positive lower bound was proved by
Hislop and M\"uller \cite{[HM]} under the assumption that the
probability density satisfies
$\rho \geq \rho_{min} > 0$. Suppose the deterministic spectrum
of $H_\omega$ is $[ \Sigma_- , \Sigma_+]$. Then, for all $\epsilon > 0$,
there is a constant $ C_\epsilon >0$, depending on $\rho_{min}$,
such that $n(E) > C_\epsilon$
for all $E \in [ \Sigma_- + \epsilon, \Sigma_+ - \epsilon ]$.
Exponential decay of fractional moments of Green's functions for
random Schr\"odinger operators was established in certain energy
regimes by Aizenman and Molchanov \cite{[AM]}, by Aizenman
\cite{[Aizenman]}, and by Aizenman, Schenker, Friedrich and
Hundertmark \cite{[ASFH]}.

Additionally, Minami's proof rests on a certain correlation estimate
for the second moment of the resolvent. It is this estimate that
interests us here. We present a new proof of this estimate that
generalizes Minami's result in two ways: 1) it holds for general
bounded, selfadjoint Hamiltonians $H_0$, including magnetic
Schr\"odinger operators and operators with decaying, off-diagonal
matrix elements, and 2) it holds for higher-order moments of the
Green's function. These generalizations of Minami's estimate were
recently also obtained by Graf and Vaghi \cite{[GV]} with a
different method, which we outline in section~\ref{sec:GV} below.

We also apply this moment bound to prove a new estimate on the
probability that there are at least $n$ eigenvalues of a local
Hamiltonian in a given energy interval. We interpret this as an
$n$-{\it level Wegner-type estimate} that bounds the probability of
$n$ eigenvalues being in the same energy interval. As such, it is a
measure of the correlation between multiple eigenvalues.

Minami's estimate may be stated in several ways. For $z \in \CM^+$,
we let $R_\Lambda (z) = (H_\Lambda - z)^{-1}$ denote the resolvent of the
finite-volume Hamiltonian on $\ell^2 (\Lambda)$. The corresponding Green's
function is denoted by $G_\Lambda ( x, y ;z)$, for $x,y \in \Lambda$. Minami
stated the estimate this way (Lemma 2, \cite{Mi96}).

\begin{lemma}\label{minami1}
For any $z \in \CM^+$, any cube $\Lambda \subset \ZM^d$, and for any $x, y
\in \Lambda$ with $x \neq y$, we have
\beq\label{minamiest1}
\EM \left[ \det \left( \begin{array}{ll}
                \Im G_\Lambda (x,x ;z)  & \Im G _\Lambda ( x,y ;z) \\
                  \Im G_\Lambda ( y,x; z) & \Im G_\Lambda ( y, y; z)
                  \end{array}
            \right) \right] \leq \pi^2 \| \rho \|_\infty^2 .
\eeq
\end{lemma}

However, for many purposes, it is clearer to note that terms as on
the left side of (\ref{minamiest1}) arise when evaluating $\left( Tr
\{\Im R_\Lambda\}\right)^2 - Tr \{ (\Im R_\Lambda)^2\}$ in the
canonical basis of $\ell^2(\Lambda)$. Thus (\ref{minamiest1})
produces the bound
\beq\label{minamiest2} \EM \Big[ \left( Tr \{\Im
R_\Lambda (z)\}\right)^2 -
Tr \{ (\Im R_\Lambda (z) )^2\} \Big] \leq \pi^2
\| \rho \|_\infty^2 | \Lambda |^2.
\eeq
As written in the appendix of \cite{[KM]},
estimate (\ref{minamiest2}) easily leads to the bound
\beq\label{minamiest3}
\EM \{ ( Tr E_\Lambda (J) )^2 - Tr E_\Lambda
(J) \} \leq \pi^2 \| \rho \|_\infty^2 | J |^2 | \Lambda|^2 ,
\eeq
for any interval $J \subset \RM$.
This estimate (\ref{minamiest3}) was used by Klein and Molchanov \cite{[KM]}
to provide a new proof of
the simplicity of eigenvalues for random Schr\"odinger operators on
the lattice, previously shown by Simon \cite{[Simon]} with other
methods. In fact, from Chebyshev's inequality, we can write
(\ref{minamiest3}) as
\beq\label{km01} \PM \{ Tr E_\Lambda ( J)
\geq 2 \} \leq \frac{\pi^2}{2}\; \| \rho \|_\infty^2 | J|^2 |
\Lambda|^2.
\eeq
Note, for comparison, that the Wegner estimate
states that
\beq\label{wegner1} \PM \{ Tr E_\Lambda ( J) \geq 1 \}
\leq \pi \| \rho \|_\infty | J | |\Lambda|.
\eeq
It is crucial for
the applications that in the bound of the left side of (\ref{km01})
the exponents of both the volume factor and the length of the
interval $|J|$ be greater than one. We mention that a bound of the
type (\ref{km01}) is not known for random Schr\"odinger operators on
$L^2 ( \Lambda)$, for $\Lambda \subset \RM^d$. This is the
main remaining obstacle to extending Minami's result on Poisson
statistics for energy level spacings, and the
Klein-Molchanov proof of the simplicity of eigenvalues, for energies
in the strong
localization regime, to
continuum Anderson-type models.
Our original motivation for this work was two-fold: First, we wanted to
find another proof of Minami's miracle in order to better understand it,
and, secondly, we wanted to try to generalize the Minami estimate so that
it was applicable to other
models.

\subsection{Contents of this Article}

We state our main results, a generalization of Minami's correlation
estimate, Theorem \ref{minami05.th-minami}, and its application to
an $n$-level Wegner estimate, Theorem \ref{minami05.th-locdet}, in
section 2. Given Theorem \ref{minami05.th-minami}, we prove the
$n$-level Wegner estimate in section 2. We prove the main
correlation estimate in section 3 using a Gaussian integral
representation of the determinant. In section 4, we discuss
applications to energy level statistics and the simplicity of
eigenvalues in the localization regime for general Hamiltonians, and
a related proof of the correlation estimate due to Graf and Vaghi
\cite{[GV]}. We also discuss the related works by Nakano \cite{[N]}
and Killip and Nakano \cite{[KN]} on joint energy-space
distributions. For convenience, a proof of the Schur complement
formula is presented in the appendix.

\section{The Main Results}
\label{minami05.sec-results}

We consider random perturbations of a fixed, bounded, selfadjoint,
background operator $H_0$.
The general Anderson model is given by the following random Hamiltonian
acting on $\ell^2(\ZM^d)$,

\begin{equation}
\label{genand1} 
H_{\omega}\psi(x) \;=\; H_0 \psi (x) \;+\;
   V (x)\psi(x), ~~~x \in \ZM^d,
\hspace{2cm}
  \psi \in \ell^2(\ZM^d)\,,
\end{equation}

\noindent that generalizes
(\ref{minami05.eq-andersonmodel00}).
The potential
$\omega = \left( V (x)\right)_{x\in\ZM^d}$ is a family of independent,
identically distributed random variables $V(x)$ with distribution given by a
density $\rho(V(0))$
such that $\|\rho\|_{\infty} = \sup_{V(0)} \rho(V(0)) < \infty$.

Among the general operators $H_0$, we note the following important
examples. The first family of examples include nonrandom perturbations
of the lattice Laplacian $L$, defined by,
\beq\label{laplacian1}
(L \psi) (x ) = \sum_{y;|y-x|=1} \;
  \psi(y),
\eeq
by $\Gamma \subset \ZM^d$-periodic potentials $V_0$ on $\ZM^d$ so that $H_0
= L + V_0$. Here, the group $\Gamma$ is
some nondegenerate subgroup of $\ZM^d$ like $N \ZM^d$, for some $N > 1$.
The second
family of examples consists of bounded, selfadjoint, operators $H_0$ with
exponentially-decaying, off-diagonal matrix elements.
The third family of examples
are discrete Schr\"odinger operators with magnetic fields,
\beq\label{magnetic1}
(H_0 \psi) (x ) = \sum_{y;|y-x|=1} ( \psi (x) - e^{i A(x,y)} \psi (y) ),
\eeq
where $A(x,y) = - A(y,x)$ is nonvanishing for $|x-y| =1$ and takes values
on the torus.
The operator $H_0$ need not be a Schr\"odinger operator but simply a bounded
selfadjoint operator for
Theorems \ref{minami05.th-minami}, \ref{minami05.th-locdet}, and
\ref{minami05.th-wegner}.
The boundedness of $H_0$
is not essential, but we will require this in order to avoid
selfadjointness problems.

When we consider localization and eigenvalue level
spacing statistics in section 4,
we will require that, in addition, the selfadjoint operator
$H_0$ is translation invariant with
the off-diagonal
matrix elements, $| \langle x | H_0 | y \rangle |$, decaying
sufficiently fast in $| x - y|$. We
will discuss the required properties of $H_0$ further in section
\ref{applications}.

\vspace{.2cm}

\subsection{Generalization of Minami's Correlation
Estimate}\label{correlest1}

\noindent For any subset $\Lambda \subset \ZM^d$, we define
$P_\Lambda$ to be the orthogonal projection onto $\ell^2(\Lambda)$,
so that $P_\Lambda f(x) = \sum_{y \in \Lambda} f(y) \delta_{x,y}$.
By $H_{\Lambda}$ we denote the restriction $P_{\Lambda} H_\omega
P_{\Lambda}$ of $H_\omega$ to $\ell^2(\Lambda)$.

\noindent Let $\Delta \subset \Lambda$ and note that $V$ commutes
with $P_\Delta$: $V_\Delta = P_\Delta V=VP_\Delta$.
For $z\in \CM$, with $\Im (z) >0$,
the matrix-valued function
$g_\Delta(z) = P_\Delta (H_{\Lambda}-z)^{-1}P_\Delta$
has the following property (see \cite{Mi96} for the case $n=2$).

\begin{theo}
\label{minami05.th-minami}
For $\Im(z) >0$ and any subset
$\Delta \subset \Lambda$, with $|\Delta|=n$,
the following inequality holds:
\beq\label{minami01}
\EM \left(    \det\{\Im g_\Delta(z)\}
      \right) \leq \pi^n\|\rho\|_{\infty}^n\,,
\hspace{2cm} |\Delta|=n\,.
\eeq
\end{theo}

\noindent A new proof of this result, using the representation of
the square root of a determinant by a Gaussian integral, will be
given in section~\ref{minami05.sect-main}. It is a generalization of
Minami's result Lemma~\ref{minami1} where $H_0$ is the discrete
Laplacian $L$, defined in (\ref{laplacian1}),
and $n = 2$. In the case $n =2$, we may write $P_\Delta =
| x \rangle \langle x| + |y \rangle \langle y |$, for $x \neq y$, so
that \beq\label{minami2} g_\Delta ( z) = \left(
\begin{array}{cc}
                      G_\Lambda (x,x; z) & G_\Lambda (x,y; z) \\
                          G_\Lambda (y,x;z) & G_\Lambda (y,y;z)
                         \end{array}
           \right)
\eeq where, as above, $G_\Lambda (x,y;z)$ is the Green's function
for $H_\Lambda$. Thus Lemma~\ref{minami1} follows from
Theorem~\ref{minami05.th-minami} if we note that in this case one
has $G_{\Lambda}(x,y;z)=G_{\Lambda}(y,x;z)$ and thus $\Im g_\Delta =
(g_\Delta-g_\Delta^*)/2\imath$ in (\ref{minami01}) is the same as
the matrix on the left of (\ref{minamiest1}).

\subsection{The $n$-level Wegner Estimate}\label{correlest2}

We use Theorem \ref{minami05.th-minami}
to prove a new estimate about multiple eigenvalue correlations. We begin
with
the observation
that the left hand side of (\ref{minami01}) can be interpreted in terms of
the
eigenvalues of an operator acting on a certain antisymmetric subspace of a
finite tensor product.

\begin{theo}
\label{minami05.th-locdet} Let $A=A^\ast$ be a selfadjoint operator
on $\ell^2(\Lambda)$, $\Lambda \subset \ZM^d$ finite, with
eigenvalues $a_1 \leq a_2 \leq \cdots \leq a_N$, where
$N=|\Lambda|$. Then, the following holds

$$\sum_{\Delta\subset \Lambda;|\Delta|=n}
   \det\{P_\Delta A P_\Delta\} \;=\;
    \sum_{1\leq i_1<\cdots < i_n \leq N}
     a_{i_1}\dots a_{i_n}\,.
$$

\noindent Moreover, if $\Hh_n = \ell^2(\Lambda)^{\wedge n}$ is the
{\em $n$-fermion} subspace, let $A^{\wedge n}$ be the restriction of
$A^{\otimes n}$ to $\Hh_n$. Then
$$\sum_{\Delta\subset \Lambda;|\Delta|=n}
   \det\{P_\Delta A P_\Delta\} \;=\;
    \TR_{\Hh_n}\left(A^{\wedge n}\right)\,.
$$
\end{theo}

\noindent {\bf Proof: } The first identity is a trivial consequence
of the second. For indeed, the eigenvalues of $A^{\wedge n}$ are products
of the form $a_{i_1}\dots a_{i_n}$ with $1\leq i_1<\cdots < i_n \leq N$
and the trace is the sum of the eigenvalues.

\vspace{.2cm}

\noindent To prove the second identity, for each $x\in\Lambda$,
let $e_x$ be the unit vector in $\Hh_1=\ell^2(\Lambda)$ supported by $x$,
namely $e_x(y) = \delta_{x,y}$. Then $\{e_x\,;\, x\in \Lambda\}$
is an orthonormal basis of $\Hh_1$. Let $\Lambda$ be ordered so that we
may write $x_1 < x_2 <
\cdots < x_N$. Then,
$\{ e_{x_{i_1}}\wedge \cdots \wedge e_{x_{i_n}} ~| ~x_{i_j} \in \Lambda,
~1 \leq i_j \leq N \}$ is
an orthonormal basis of $\Hh_n$ if we restrict to
indices so that $x_{i_1} < x_{i_2} < \cdots < x_{i_n}$, with
$1 \leq i_j \leq N$. Thus, the trace on $\Hh_n$ can be expanded as

$$\TR_{\Hh_n}\left(A^{\wedge n}\right)\;=\;
   \sum_{x_1 <\cdots <x_n}
    \langle e_{x_1}\wedge \cdots \wedge e_{x_n}|
            A^{\wedge n}
            e_{x_1}\wedge \cdots \wedge e_{x_n}
    \rangle\,.
$$

\noindent If $\Delta = \{x_1, \cdots, x_n\}$ (where the labeling is such
that $x_1 <x_2<\cdots <x_n$), then the definition of the determinant gives
\begin{eqnarray*}
\langle e_{x_1}\wedge \cdots \wedge e_{x_n}|
            A^{\wedge n}
            e_{x_1}\wedge \cdots \wedge e_{x_n}
  \rangle  &=&
   \langle e_{x_1}\wedge \cdots \wedge e_{x_n}|
            (P_\Delta A P_\Delta)^{\wedge n}
            e_{x_1}\wedge \cdots \wedge e_{x_n}
    \rangle \\
  &=& \det\{P_\Delta A P_\Delta\} \,.
\end{eqnarray*}
\hfill $\Box$

We can now combine the previous two Theorems to generalize
(\ref{km01}) and prove the following $n$-level Wegner estimate.
We point out that this estimate holds for all energy intervals $J$
in the spectrum of $H_\Lambda$.

\begin{theo}
\label{minami05.th-wegner} For any positive integer $n$, and
interval $J\subset \RM$, and any cube $\Lambda \subset \ZM^d$ we
have \beq\label{generalwegner} \PM(\TR E_{\Lambda}(J) \ge n) \le
\frac{\pi^n}{n!} \|\rho\|^n_{\infty} |J|^n |\Lambda|^n. \eeq
\end{theo}

\noindent {\bf Proof:} By taking $A= \Im R_{\Lambda}(z)$, $\Im z>0$,
in Theorem~\ref{minami05.th-locdet} and using the result from
Theorem~\ref{minami05.th-minami} we get
\begin{eqnarray} \label{generalminami1}
\EM \left( \TR_{\Hh_n} ((\Im R_{\Lambda}(z))^{\wedge n})\right) & =
& \sum_{\Delta\subset \Lambda;\,|\Delta|=n} \EM(\det \{\Im
g_{\Delta}(z)\}) \nonumber \\ & \le & {|\Lambda| \choose n} \pi^n
\|\rho\|_{\infty}^n.
\end{eqnarray}
For $\zeta=\sigma+i\tau$, $\sigma \in \RM$, $\tau>0$, define
\[ f_{\zeta}(x) = \frac{\tau}{(x-\sigma)^2+\tau^2}.\]
If $(a,b)\subset J \subset [a,b]$ and $z:=(a+b+i|J|)/2$, then
$\chi_J(x) \le |J| f_z(x)$ for all $x\in \RM$ and thus, by the
spectral theorem,
\[ E_{\Lambda}(J) \le |J| \Im R_{\Lambda}(z). \]
This carries over to $\Hh_n$ as
\begin{equation} \label{fermionbound}
E_{\Lambda}(J)^{\wedge n} \le |J|^n (\Im R_{\Lambda}(z))^{\wedge n}.
\end{equation}

\noindent If $X$ is the range of $E_{\Lambda}(J)$, then
$E_{\Lambda}(J)^{\wedge n}$ is the orthogonal projection onto the
subspace $X^{\wedge n}$ of $\Hh_n$. Thus
\[ \TR_{\Hh_n} (E_{\Lambda}(J)^{\wedge n}) = \left\{
\begin{array}{ll} {\TR E_{\Lambda}(J) \choose n} & \mbox{if $\TR
E_{\Lambda}(J) \ge n$}, \\ 0 & \mbox{if $\TR E_{\Lambda}(J)<n$}.
\end{array} \right. \]

\noindent Chebyshev's inequality and the elementary fact that $k/n
\le {k \choose n}$ for all $k\ge n$ imply that
\begin{eqnarray*}
\PM( \TR E_{\Lambda}(J)\ge n) & \le & \frac{1}{n} \EM \left( (\TR
E_{\Lambda}(J)) \cdot \chi_{\{\TR E_{\Lambda}(J) \ge n\}} \right) \\
& \le & \EM \left( {\TR E_{\Lambda}(J) \choose n} \cdot \chi_{\{\TR
E_{\Lambda}(J) \ge n\}} \right) \\ & = & \EM (\TR_{\Hh_n}
E_{\Lambda}(J)^{\wedge n}).
\end{eqnarray*}
Using the bounds (\ref{fermionbound}) and (\ref{generalminami1})
finally yields the desired result,
\begin{eqnarray*}
\PM (\TR E_{\Lambda}(J) \ge n) & \le & |J|^n \EM \left( \TR_{\Hh_n}
(\Im R_{\Lambda}(z))^{\wedge n} \right) \\
& \le & |J|^n {|\Lambda| \choose n} \pi^n \|\rho\|_{\infty}^n \\ &
\le & \frac{\pi^n}{n!} \|\rho\|_{\infty}^n |J|^n |\Lambda|^n.
\end{eqnarray*}
\hfill $\Box$


  \section{The Generalized Minami Correlation Estimate}
  \label{minami05.sect-main}

\noindent Our approach to the Minami correlation estimate Lemma
\ref{minami1}, and its generalization, is to work with the
resolvent, rather than the Green's function. We use the Schur
complement formula to isolate the random variables and the
representation of the inverse of the square root of a determinant by
a Gaussian integral, see (\ref{gaussrep1}). The proof of
Theorem~\ref{minami05.th-minami} requires several steps. As in
section \ref{minami05.sec-results}, we let $\Delta \subset \Lambda$
and denote the orthonormal projection onto $\ell^2(\Delta)$ by
$P_\Delta$. We define $\tilde{H}_\Lambda = H_\Lambda - V_\Delta$,
and for $z\in \CM$, with $\Im (z) >0$, we define the matrix-valued
functions $\tilde{g}_\Delta(z) = P_\Delta
(\tilde{H}_{\Lambda}-z)^{-1}P_\Delta$ and $g_\Delta(z) = P_\Delta
(H_{\Lambda}-z)^{-1}P_\Delta$.

  \subsection{Schur's Complement and Kre\u{\i}n's Formula}
  \label{minami05.ssect-}

\begin{lemma}
\label{minami05.lem-krein}
The following formula holds

$$g_\Delta(z) = \frac{1}{V_\Delta + \tilde{g}_\Delta(z)^{-1}}
\hspace{2cm}\mbox{\small\bf Kre\u{\i}n's formula}
$$
\end{lemma}

\noindent  {\bf Proof: } The {\em Schur complement formula} \cite{Schur17}
(also called Feshbach's
projection method \cite{Feshbach58}
\footnote{
The Schur complement method \cite{Schur17} is widely used in numerical
analysis under this name, while Mathematical Physicists prefer the reference
to
Feshbach\cite{Feshbach58}. It is also called Feshbach-Fano \cite{Fano35} or
Feshbach-L\"owdin \cite{Lowdin62} in
Quantum Chemistry. This method is used in various algorithms in Quantum
Chemistry ({\em ab initio} calculations),
in Solid State Physics (the muffin tin approximation, LMTO) as well as in
Nuclear Physics.
The formula used above is found in the original paper of Schur
\cite{Schur17} (the formula is on p.217).
The formula has been proposed also by an astronomer Tadeusz Banachiewicz in
1937, even though closely
related results were obtained in 1923 by Hans Boltz and in 1933 by Ralf
Rohan \cite{PS04}.
Applied to the Green function of a selfadjoint operator with finite rank
perturbation,
it becomes the Kre\u{\i}n formula \cite{Krein46}.
})
\,states
that if $H=H^\ast$ is a bounded, selfadjoint operator on some
Hilbert space and if $P=\id -Q$ is an orthonormal projection, then

\beq\label{schur1}
P\frac{1}{H-z}P\;=\; \frac{1}{H_{eff}(z) - z P }\,,
\hspace{1cm} H_{eff}(z) = PHP + PHQ\frac{1}{QHQ-z}QHP\,.
\eeq

\noindent For completeness we provide a proof of (\ref{schur1}) in
section~\ref{sec:schur}. Applied to $g_\Delta(z) = P_\Delta
(H_{\Lambda}-z)^{-1}P_\Delta$ gives

$$g_\Delta(z)^{-1}\;=\;
   H_\Delta + P_\Delta H_\Lambda P_{\Lambda\setminus \Delta}
              \frac{1}{H_{\Lambda\setminus \Delta}-z}
              P_{\Lambda\setminus \Delta}H_\Lambda P_\Delta\,.
$$

\noindent By definition, $H_\Delta =\tilde{H}_\Delta + V_\Delta$ while
$H_{\Lambda\setminus \Delta}=\tilde{H}_{\Lambda\setminus \Delta}$, so that,
applying the Schur complement formula to $\tilde{g}_\Delta(z)$ instead,
gives the desired result

$$g_\Delta(z)^{-1} \;=\; V_\Delta + \tilde{g}_\Delta(z)^{-1}\,.
$$
\hfill $\Box$

\begin{lemma}
\label{minami05.lem-pos} If $\Im z>0$, then $\Im g_\Delta(z) >0$ and
$-\Im \{\tilde{g}_\Delta(z)^{-1}\}>0$.
\end{lemma}

\noindent  {\bf Proof: } The resolvent equation gives

$$\Im g_\Delta(z) = P_\Delta \frac{\Im z}{|H_\Lambda-z|^2} P_\Delta
      \;>\;0\,,
\hspace{2cm}
\Im \tilde{g}_\Delta(z) = P_\Delta \frac{\Im z}{|\tilde{H}_\Lambda-z|^2}
P_\Delta
      \;>\;0\,.
$$

\noindent Using $A^{-1}-A^{-1\,\ast}= A^{-1}\{A^\ast - A\}A^{-1\,\ast}$
gives the other inequality.
\hfill $\Box$

\begin{lemma}
\label{minami05.lem-det}
The following formula holds:

$$\det\{\Im g_\Delta(z)\}\;=\;
   \frac{\det\{-\Im \tilde{g}_\Delta(z)^{-1}\}}
     {|\det\{V_\Delta +\tilde{g}_\Delta(z)^{-1}\}|^2}
$$
\end{lemma}

\noindent  {\bf Proof: } By definition of the imaginary part, using
Lemma~\ref{minami05.lem-krein} gives

\begin{eqnarray*}
\Im g_\Delta(z) &=& \frac{g_\Delta(z)-g_\Delta(z)^\ast}{2\imath}\\
&=& \frac{1}{V_\Delta +\tilde{g}_\Delta(z)^{-1}}
\left(
   \frac{\tilde{g}_\Delta(z)^{-1\,\ast}-\tilde{g}_\Delta(z)^{-1}}{2\imath}
\right)
    \frac{1}{V_\Delta +\tilde{g}_\Delta(z)^{-1\,\ast}}\\
&=& -\frac{1}{V_\Delta +\tilde{g}_\Delta(z)^{-1}}
      \left[\Im \tilde{g}_\Delta(z)^{-1}\right]
    \frac{1}{V_\Delta +\tilde{g}_\Delta(z)^{-1\,\ast}}.
\end{eqnarray*}

\noindent Taking the determinant of both sides gives the result.
\hfill $\Box$

\begin{coro}
\label{minami05.cor-minaineq1} If $\EM_S$ denotes the average over
the potentials $V_x$ for $x\in S$, the following estimate holds:
$$\EM_\Lambda\left(
  \det\{\Im g_\Delta(z)\}
   \right)\;\leq\;
\EM_{\Lambda\setminus \Delta}\left(
  \|\rho\|_{\infty}^n \det\{-\Im \tilde{g}_\Delta(z)^{-1}\}
   \int_{\RM^\Delta}
    dV_\Delta
     \frac{1}{|\det\{V_\Delta +\tilde{g}_\Delta(z)^{-1}\}|^2}
   \right)
$$
\end{coro}

\noindent  {\bf Proof: } By definition and since $|\Delta|=n$, if
$f$ is a nonnegative function of $V=(V_x)_{x\in \Lambda}$ then

$$\EM_\Lambda (f) \;=\;
   \int_{\RM^\Lambda}
    \prod_{x\in\Lambda} dV_x \;\rho(V_x) \;f(V) \; \leq \;
     \|\rho\|_{\infty}^n
      \EM_{\Lambda\setminus \Delta}\left(
       \int_{\RM^\Delta}
        \prod_{x\in\Delta} dV_x f(V)
                                   \right).
$$

\noindent Since $\tilde{g}_\Delta(z)$ does not depend on $V_\Delta$, the
result follows from Lemma~\ref{minami05.lem-det}.
\hfill $\Box$

\begin{lemma}
\label{minami05.lem-gauss}
Let $M$ be a complex $n\times n$ matrix such that $M=B-\imath A$
with $B=B^\ast$ and $A$ positive definite. Then, taking the principal branch
of the square root,

\beq\label{gaussrep1}
\frac{1}{\sqrt{\det{M}}} \;=\; e^{\imath n \pi/4}
   \int_{\RM^n} \frac{d^n u}{(2\pi)^{n/2}}\;
    e^{-\imath \langle u| Mu\rangle/2}
\eeq
\end{lemma}

\noindent  {\bf Proof: } Since $A >0$, it follows that $\imath M$ has a
positive definite real part, so that the integral converges and is analytic
in $M$. The formula follows from standard Gaussian integrals.
\hfill $\Box$

\begin{lemma}
\label{minami05.lem-sphe}
Let $F$ be an integrable function on $\RM^2\times \RM^2$. Then the following
formula holds

\beq\label{integral1}
\int_{\RM^2\times \RM^2} d^2\uV\; d^2\vV\;\;
   F(\uV,\vV)\;
    \delta\left(
                \frac{\uV^2-\vV^2}{2}
          \right) \;=\;
    \int_{\RM^2} d^2\uV
     \int_0^{2\pi} d\theta\;
      F(\uV, R_\theta \uV)\,,
\eeq

\noindent where $R_\theta$ denotes the rotation of angle $\theta$ in
$\RM^2$.
\end{lemma}

\noindent  {\bf Proof: } Let $\uV$ be expressed in polar coordinates
$(r,\phi)$. The change of variable $s= \uV^2/2= r^2/2$ gives
$\uV=(\sqrt{2s},\phi)$ and  $d^2\uV=ds d\phi$. In much the same way,
$\vV$ can be expressed as $(\sqrt{2t},\psi)$. Thus the integral
becomes

\begin{eqnarray*}
& &\int_{\RM^2\times \RM^2} d^2\uV\; d^2\vV\;\;
   F(\uV,\vV)\;
    \delta\left(
                \frac{\uV^2-\vV^2}{2}
          \right) \\
& = & \int_0^\infty ds \int_0^\infty dt
     \int_0^{2\pi} d\phi \int_0^{2\pi} d\psi\;
      F(\sqrt{2s},\phi;\sqrt{2t},\psi)
      \delta(s-t) \\
& = & \int_0^\infty ds
     \int_0^{2\pi} d\phi \int_0^{2\pi} d\psi\;
      F(\sqrt{2s},\phi;\sqrt{2s},\psi)\,.
\end{eqnarray*}

\noindent Setting $\psi= \theta + \phi$ gives the result.
\hfill $\Box$

  \subsection{Proof of Theorem~\ref{minami05.th-minami}}
  \label{minami05.ssect-proofth}

\noindent Thanks to Corollary~\ref{minami05.cor-minaineq1}, the main
Theorem~\ref{minami05.th-minami} follows from the following inequality.

\begin{lemma}
\label{minami05.lem-minaineq2}
The following estimate holds

$$J\;:=\;
   \int_{\RM^\Delta}
    dV_\Delta
     \frac{1}{|\det\{V_\Delta +\tilde{g}_\Delta(z)^{-1}\}|^2}
     \;\leq\;
    \frac{\pi^n}{\det\{-\Im\tilde{g}_\Delta(z)^{-1}\}}\,.
$$
\end{lemma}

\noindent  {\bf Proof: }
Using the Gaussian integral in
Lemma~\ref{minami05.lem-gauss}, the integral $J$ can be written as

\begin{equation}
\label{minami05.eq-J1}
J=
   \int_{\RM^\Delta}
    dV_\Delta
   \int_{(\RM^\Delta)^{\times 4}}
    \frac{d^n u_1 d^n u_2}{(2\pi)^n}
    \frac{d^n v_1 d^n v_2}{(2\pi)^n}
e^{-1/2\sum_{i=1,2}
  \{ \langle u_i |\imath (V_\Delta +\tilde{g}_\Delta(z)^{-1})u_i\rangle -
\langle v_i |\imath (V_\Delta
+\tilde{g}_\Delta(\overline{z})^{-1})v_i\rangle \}
   }\,.
\end{equation}

\noindent Let $\uV(x) = (u_1(x), u_2(x))\in \RM^2$ where
$u_i=\left(u_i(x)\right)_{x\in\Delta} \in \RM^\Delta$. In much the same way
let $\vV(x) = (v_1(x), v_2(x))\in \RM^2$ be used in this integral. The term
$V_x$ appears in the Gaussian exponent with the factor $(-\imath/2)
V_x(\uV(x)^2-\vV(x)^2)$. Hence integration over $V_x$ gives

$$\int_\RM dV_x \;
   e^{
   -\imath V_x (\uV(x)^2-\vV(x)^2)/2
      } \;=\;
  2\pi \delta\left(
                \frac{\uV(x)^2-\vV(x)^2}{2}
          \right)\,.
$$

\noindent Inserting this result in eq.~(\ref{minami05.eq-J1}), using
Lemma~\ref{minami05.lem-sphe} leads to

$$J= \prod_{x\in \Delta}\int_0^{2\pi} d\theta_x\;
    \int_{(\RM^\Delta)^{\times 2}}
    \frac{d^{2n} \uV }{(2\pi)^n}
e^{-1/2 \{
   \langle \uV |\imath \tilde{g}_\Delta(z)^{-1}\uV\rangle +
     \langle R(\theta)\uV |\imath \tilde{g}_\Delta(\overline{z})^{-1}
R(\theta)\uV\rangle} \} \,,
$$

\noindent where $R(\theta)$ is the orthogonal $2n\times 2n$ matrix acting on
$\uV=(\uV(x))_{x\in\Delta}$ by

$$\left( R(\theta)\uV \right) (x) \;=\;
     R_{\theta_x} \uV(x)\,.
$$

\noindent
Because of
Lemma~\ref{minami05.lem-pos}, we know
that $-\Im \tilde{g}_\Delta(z)^{-1} >0$,
so that the Gaussian term can be bounded from above by

$$J \leq \prod_{x\in \Delta}\int_0^{2\pi} d\theta_x\;
    \int_{(\RM^\Delta)^{\times 2}}
    \frac{d^{2n} \uV }{(2\pi)^n}
e^{-1/2 \{
   \langle \uV | (-\Im \tilde{g}_\Delta(z)^{-1} ) \uV\rangle +
     \langle R(\theta)\uV | (-\Im \tilde{g}_\Delta(z)^{-1} )
R(\theta)\uV\rangle}\,.
$$

\noindent Thus a Schwarz inequality, the rotational invariance of
the measure $d^{2n} \uV$ and another use of the Gaussian formula
given in Lemma~\ref{minami05.lem-gauss} gives the bound
$$J \leq \prod_{x\in \Delta}\int_0^{2\pi} d\theta_x\;
    \int_{(\RM^\Delta)^{\times 2}}
    \frac{d^{2n} \uV }{(2\pi)^n}
e^{-
   \langle \uV | ( -\Im \tilde{g}_\Delta(z)^{-1} ) \uV\rangle}\;=\;
\frac{\pi^n}{\det\{-\Im \tilde{g}_\Delta(z)^{-1}\}}\,,
$$
proving the theorem.
\hfill $\Box$


\section{Applications of the Correlation Estimate and Related Results}
\label{applications}

\subsection{Level Statistics and Simplicity of Eigenvalues}

The proof of Poisson statistics for the eigenvalue level spacing
in the thermodynamic limit
for general Anderson Hamiltonians
$H_0 + V$ as in (\ref{genand1}) follows as in Minami's article provided
several other conditions are satisfied.
In addition to the selfadjointness and boundedness of $H_0$,
we require that $H_0$ be translation invariant, so that the DOS exists, and
that the off-diagonal matrix elements of $H_0$ decay exponentially in
$|x-y|$ with a uniform rate.
In addition to the
positivity of the DOS at energy $E$, discussed in the introduction,
Minami requires the exponential
decay of the expectation of a fractional moment of the Green's
function of $H_\omega$.
Let us describe this fractional moment condition. The
Green's function $G_\Lambda (x,y; z)$ for the restriction of $H_\omega$ to
a finite cube $\Lambda
\subset \ZM^d$ with Dirichlet boundary conditions
is required to satisfy the following bound. There is
some $0 < s < 1$ and constants $C_s > 0$ and $\alpha_E > 0$ so that
\beq\label{expect1} \EM \{ | G_\Lambda ( x, y; E+i \epsilon ) |^s \}
\leq C_s e^{-\alpha_E | x - y|}, \eeq provided $x \in \Lambda$, $y
\in \partial \Lambda$, and $z \in \{ w \in \CM ~| ~\Im w > 0, | w -
E | < r \}$, for some $r>0$.

\begin{coro}\label{minamitheorem2}
Consider the general Anderson model (\ref{genand1}) with a bounded,
translation-invariant, selfadjoint $H_0$ having matrix elements
satisfying $|\langle x|H_0|y\rangle| \le Ce^{-\eta|x-y|}$ for some
$C<\infty$ and $\eta>0$.
Suppose that the
DOS $n(E)$ for $H_\omega$
exists at energy $E$ and is positive. Suppose also that
the expectation of some fractional moment of the Green's function
of $H_\omega$
decays exponentially fast as described in (\ref{expect1}).
Then, the point process (\ref{poisson1}) converges
weakly to the Poisson point process with intensity measure $n(E)
~dx$.
\end{coro}

\noindent We don't provide a detailed proof of this here, as it is
easily checked that under the assumption of exponential decay of
$|\langle x|H_0|y\rangle|$  the remaining arguments of Minami's
proof of Poisson statistics go through. Translation invariance of
$H_0$ comes in as an extra assumption to guarantee ergodicity of
$H_{\omega}$, and thus existence of the IDS (\ref{ids1}).

Exponential decay of $|\langle x|H_0|y\rangle|$ implies the strong
localization condition (\ref{expect1}) at extreme energies or high
disorder \cite{[AM]}, or, for low disorder, at band edges
\cite{[Aizenman]}.
Also, exponential bounds of the form (\ref{expect1}) imply almost
sure pure point spectrum for the energies at which they hold and
exponential decay of the corresponding eigenfunctions.

These conditions on $H_0$ and the decay estimate (\ref{expect1})
also insure that the result of Klein and Molchanov \cite{[KM]}
(which uses \cite{[AM]} and thus rapid off-diagonal decay of the
matrix elements of $H_0$) on the almost sure simplicity of
eigenvalues are applicable in the above situation, thus:

\begin{coro}\label{minamitheorem3}
The eigenvalues of the general Anderson model considered in
Corollary~\ref{minamitheorem2} in the region of localization are
simple almost surely.
\end{coro}

In fact, all of the above can be extended to $H_0$ with sufficiently
rapid power decay of the off-diagonal elements. The works
\cite{[AM]} and \cite{[Aizenman]} discuss how a result somewhat
weaker than (\ref{expect1}) can be obtained in this case. In
particular, this only gives power decay of eigenfunctions, which for
sufficiently fast power decay still allows to apply the result of
\cite{[KM]}. Moreover,
a thorough analysis of Minami's proof shows
that it works for suitable power decay.

\subsection{A different proof of Theorem 2} \label{sec:GV}

After we finished the proof of Theorem \ref{minami05.th-minami}, we
received the preprint of Graf and Vaghi \cite{[GV]} in which they
proved essentially the same result using a different approach. One
of their main motivations was to eliminate Minami's symmetry
condition on the Green's function $G_\Lambda ( x,y;z) = G_\Lambda
(y,x;z)$ thus allowing magnetic fields as in (\ref{magnetic1}). They
base their calculation on the following lemma. By $\mbox{diag} ~(
v_1, \ldots, v_n)$, we mean the $n \times n$-matrix with only
nonzero diagonal entries $v_1, \ldots, v_n$.

\begin{lemma}\label{gv1}
Let $A = (a_{ij})$ be an $n \times n$ matrix with $\Im A > 0$. then,
\beq\label{gvbound1}
\int ~dv_1 \cdots d v_n  ~\det ( \Im [ ~\mbox{diag} ~( v_1, \ldots, v_n)
- A ]^{-1}
) \leq \pi^n .
\eeq
\end{lemma}

It is not surprising that the proof of Lemma \ref{gv1} involves the
Schur complement formula. One applies Lemma \ref{gv1} by noting that
the argument of the determinant on the left side of
(\ref{minamiest1}) (for the case $n=2$) may be written as the
imaginary part of a matrix of the form $[ ~\mbox{diag} ~( v_1,
\ldots, v_n) - A ]^{-1}$ by Krein's formula, where $A$ is obtained
from $H_\Lambda$ by setting $V(x) = V(y) = 0$. Without explicitly
stating it, Graf and Vaghi also indicate that a bound as in
(\ref{minami01}) follows from (\ref{gvbound1}) for general $n$.

The key to proving Lemma \ref{gv1} for $n=2$ are the integral formulas
\beq\label{gv2}
\int_{\RM} ~dx ~\frac{1}{|ax + b|^2} = \frac{\pi}{\Im ( \overline{b}a)},
\eeq
assuming $a, b \in \CM$ and $\Im ( \overline{b} a) > 0$, and
\beq\label{gv3}
\int_{\RM} ~dx ~\frac{1}{ax^2 + bx+ c}  = \frac{ 2\pi}{\sqrt{ 4ac - b^2}},
\eeq
assuming $a, b, c \in \RM$, $a > 0$, and $4ac - b^2 > 0$.
The case for general $n$ is obtained by induction.

\subsection{Joint Energy-Space Distributions}

We mention two related results of interest. Nakano \cite{[N]}
recently obtained some quantitative results providing insight into
the idea, going back to Mott, that when eigenvalues in the
localization regime are close together, the centers of localization
are far apart. Roughly, Nakano proves that for any subinterval $J$
of energies in the localization regime with sufficiently small
length, there is at most one eigenvalue of $H_\omega$ in $J$ with a
localization center in a sufficiently large cube about any point
with probability one. His proof uses Minami's estimate in the form
(\ref{minamiest3}) and the multiscale analysis. In this sense, the
centers of localization are repulsive. On the other hand, if one
studies an appropriately scaled space and energy distribution of the
eigenfunctions in the localization regime in the thermodynamic
limit, Killip and Nakano \cite{[KN]} proved that this distribution
is Poissonian, extending Minami's work for the Anderson model
(\ref{minami05.eq-andersonmodel00}).
They define a measure $d \xi$ on $\RM^{d+1}$ by the following
functional. For $f \in C_0 ( \RM )$ and $g \in C_0 (\RM^d)$,
consider the map \beq\label{kn1} f,g \rightarrow tr ( f(H) g(\cdot)
) = \int_{\RM \times \RM^d} f(E) g(x) ~d \xi (E, x). \eeq This
measure is supported on $\Sigma \times \ZM^d$, where $\Sigma \subset
\RM$ is the deterministic spectrum of $H_\omega$.  They perform a
microscopic rescaling of $d \xi$ in both energy and space to obtain
a measure $d \xi_L$ as follows \beq\label{kn2} \int f(E,x) ~d \xi_L
(E,x) = \int f( L^d (E - E_0), x L^{-1}) ~d \xi (E, x), \eeq where
$E_0$ is a fixed energy for which the density of states $n$ exists
and is positive. In the limit $L \rightarrow \infty$, they prove
that this rescaled measure converges in distribution to a Poisson
point process on $\RM \times \RM^d$ with intensity given by $n(E_0)
dE \times dx$. This work also relies on Minami's estimate
(\ref{minamiest3}) but uses the fractional moment estimates rather
than multiscale analysis. Both of these papers treat the standard
Anderson model (\ref{minami05.eq-andersonmodel00}) so that Theorem 2
extends the results to more general lattice operators of the form
$H_0 + V_\omega$.



\section{Appendix: The Schur Complement Formula} \label{sec:schur}

We prove the Schur complement formula for a selfadjoint operator $H$
and an orthogonal projection $P$ with $Q =  1 - P$ on a Hilbert
space $\mathcal{H}$. In the case that $H$ is unbounded, we assume
that $P \mathcal{H} \subset D(H)$. Let $ z \in \CM$ and suppose that
$Q(H-z)Q$ is boundedly invertible on the range of $Q$ (as always the
case for $z\in\CM \setminus \RM$). We write $R_Q(z) =
(Q(H-z)Q)^{-1}$ for the resolvent of the reduced operator. We write
$H-z$ on $\mathcal{H}$ as the matrix \beq\label{matrixresolvent1}
H-z = \left[
\begin{array}{ll}
               P(H-z)P & P H Q \\
                Q H P  & Q(H-z)Q
                 \end{array}
                 \right]
\eeq
We introduce the triangular matrix $L$ given by
\beq\label{triangle1}
L =  \left[ \begin{array}{ll}
               P  & 0 \\
               - R_Q(z) QHP & R_Q(z)
                 \end{array}
                 \right]
\eeq The {\it Schur complement} of $Q(H-z)Q$ is defined as $S(z)
\equiv P(H-z)P - PHQ R_Q(z) QHP$. We assume that $S(z)$ is boundedly
invertible on the range of $P$ (true for $z\in \CM\setminus \RM$).
Multiplying $(H-z)$ on the right by $L$ we obtain
\beq\label{resolventmatrix2} (H-z) \cdot L = \left[
\begin{array}{ll}
                S(z) & PHQ R_Q(z) \\
                0 & Q P(H-z)P
                 \end{array}
                 \right]
\eeq
This matrix may be inverted, and multiplying the inverse $( (H-z) L)^{-1}
= L^{-1} R(z)$ on the left by $L$ gives
\beq\label{resolventmatrix3}
R(z) =  \left[ \begin{array}{ll}
               S(z)^{-1} & - S(z)^{-1} PHQ R_Q(z) \\
                - R_Q(z) QHP S(z)^{-1} & R_Q(z) + R_Q(z) QHP S(z)^{-1} Q H
                  P  R_Q(z)
                 \end{array}
                 \right]
\eeq
The formula for $PR(z)P$ readily follows from (\ref{resolventmatrix3})
since in matrix notation $P$ is block diagonal.

%
%

\end{document}